\documentclass[twocolumn,secnumarabic,amssymb, nobibnotes, aps, prl, superscriptaddress]{revtex4-2}

\usepackage{pifont}
\usepackage[english]{babel}
\usepackage[utf8]{inputenc}
\usepackage[tbtags]{mathtools}
\usepackage{relsize}
\usepackage{environ}
\usepackage{helvet}									
\usepackage[T1]{fontenc}
\usepackage{lipsum}
\usepackage{graphicx}
\usepackage{dcolumn}
\usepackage[tight]{subfigure}
\usepackage{dsfont}
\usepackage{verbatim}
\usepackage{units}
\usepackage[usenames,dvipsnames]{color}
\usepackage{bm}
\usepackage{algorithmicx}
\usepackage{algpseudocode}
\usepackage{braket}
\usepackage{mathrsfs}
\usepackage{enumerate} % roman enumeration
\usepackage[normalem]{ulem}
\usepackage{parskip}
\usepackage{xfrac}
\definecolor{myBlue}{RGB}{8,81,156}
\definecolor{myRed}{RGB}{164,16,52}
\definecolor{myGreen}{RGB}{0,90,50}
\usepackage{hyperref}
\hypersetup{colorlinks=true,linktoc=all,linkcolor=myBlue,breaklinks=true,citecolor=myBlue,urlcolor=myBlue}
\addto\captionsenglish{}

\renewcommand{\to}{\tilde{0}}

\newcommand{\half}{\dfrac{1}{2}}
\newcommand{\thalf}{{\textstyle\frac{1}{2}}}

\newcommand{\Ham}{{\mathcal{H}}}

\newcommand{\mybar}[1]{\,\overline{\!{#1}}} % overline short italic

\newcommand{\rr}[1]{{\color[RGB]{151,14,83} #1}}

\definecolor{myBlue}{RGB}{8,81,156}

\usepackage{microtype}

\begin{document}

\title{Time-dependent driving and topological protection in the fractional Josephson effect}%

% \title{Fate of topological protection in the driven quantum circuits}%

\author{Ahmed Kenawy}%
%\email{a.kenawy@fz-juelich.de}
\affiliation{Peter Gr\"{u}nberg Institut, Forschungszentrum J\"{u}lich, D-52425 J\"{u}lich, Germany}
\author{Fabian Hassler}
\affiliation{Institute for Quantum Information, RWTH Aachen University, 52056 Aachen, Germany}
\author{Roman-Pascal Riwar}
\affiliation{Peter Gr\"{u}nberg Institut, Forschungszentrum J\"{u}lich, D-52425 J\"{u}lich, Germany}

 %It customizes the line spacing

%\setlength{\parskip}{12pt}
  
%ABSTRACT
\begin{abstract}

The control of any type of quantum hardware invariably necessitates time-dependent driving. If the basis depends on the control parameter, the presence of a time-dependent control field yields an extra term in the Schrödinger equation that is often neglected. Here, we examine the effect of this term in a flux-controlled Majorana junction. We show that a time-varying flux gives rise to an electromotive force which is amplified when truncating to the junction's low-energy degrees of freedom. As a result, it compromises the robustness of the ground-state degeneracy present in the absence of the drive. The resulting flattening of the energy spectrum can be measured by a strong suppression of the dc supercurrent.

\end{abstract} 

\maketitle

\emph{Introduction.}\textemdash Topological insulators and other noninteracting systems can be classified according to their symmetries, using the ten-fold way~\cite{Zirnbauer1997,Kitaev2009,Ludwig2010,Shinsei2016}. In realistic systems, the topologically protected ground-state degeneracy can be compromised by several processes. But such processes are commonly expected to be exponentially suppressed, such as the overlap of Majorana edge modes~\cite{Leijnse2012,Marra2022}, or similarly the overlap of edge modes in topological insulators~\cite{Linder2009,Lu2010,Liu2010}. The topological protection becomes algebraic in 1D superconductors~\cite{Fidkowski2011,Cheng_2015} or when coupled to a dissipative environment~\cite{McGinley2018,McGinley2019,Cooper2020} which, among others, motivates extending the concept of topological phases to non-Hermitian Hamiltonians of open systems~\cite{Bardyn2018,Kawabata2019,Riwar2019,Lieu2020,Lieu2020PRL,Bergholtz2021,Altland2021,Deng2021,Mandal2021,Kawabata2022,Liu2022,Antonio2022,Atif2023}.

An even more basic problem is the interplay between topological protection and classical time-dependent driving. Given a Hamiltonian system that depends on a tunable control parameter~$x$, it is common to include driving parametrically\textemdash that is,~$H(x) \rightarrow H[x(t)]$\textemdash which implies that the time-dependent system inherits the symmetries and topological protection from its stationary counterpart. But if the basis of~$H$ depends on~$x$, the Schrödinger equation acquires the additional term~$-i \dot{x} U^\dagger \partial_x{U}$~\footnote{Here and in the following, we set $\hbar=1$.} where the unitary~$U(x)$ encodes the basis~\cite{Messiah}. In the adiabatic limit, this term corresponds to the Berry connection. Importantly, this term is often neglected and hence the fate of the topological protection in the presence of time-dependent driving is still largely unexplored.

In order to address this fundamental question, we study superconducting circuits. Here, the influence and microscopic origin of the term~$-i \dot{x} U^\dagger \partial_x{U}$ have recently been examined for generic superconducting circuits driven by time-dependent flux~\cite{You2019,Riwar2021,Jacob2023}, in which case this term represents an electromotive force (emf). To include the aspect of topological protection, we choose to study the basic example of Majorana fermions in~$p$-wave superconductors, which may be realized in various condensed-matter systems~\cite{FuKane,Fu2009,Nadj2013,Pientka2013,Hell2017,Pientka2017,Alicea2012,Sato2017,Flensberg2021}\textemdash for example, proximitized semiconducting nanowires~\cite{Lutchyn2010,Oreg2010}. The study of Majorana junctions~\cite{Fu2009} has seen a revival on theory side~\cite{Diminguez2017,Sau2017,Feng2018,Choi2020,Svetogorov2020,Frombach2020,Wang2022}, specifically the interplay of time-dependent driving and dissipation, along with the role of the overlap between edge modes in transport across the junction. Nonetheless, full understanding of the experimentally observed suppression of the first Shapiro step and of the Landau-Zener probability in the qubit formed by coupling two Majoranas across the weak link~\cite{Rokhinson2012,Wiedenmann2016,Bocquillon2017,Wang2018,Peter2019,Calvez2019,Rosenbach2021,Mengmeng2022,Matthias2023} is still a topic of active research.

In this letter, we study the effect of the emf on topological protection in Majorana junctions, where the control parameter is the phase bias~$\phi$ across the junction. Building on previous results valid at weak driving~\cite{Kenawy2022}, here, we account for the effect of the emf to all (relevant) orders. By deriving a low-energy theory (which requires eliminating high-energy quasiparticle states), we demonstrate that while the emf term for the full system Hamiltonian is by construction linear in the voltage~$ V = \dot{\phi} / (2e) $, its effect on the low-energy description is amplified, resulting in a highly nonlinear renormalization of the effective Majorana overlap for the driven junction. Moreover, we show that this effect strongly suppresses the supercurrent, as evidenced by the $IV$ characteristics of the driven junction. This result entails modifications of various predictions regarding time-dependent driving of Majorana junctions (e.g., Refs.~\cite{Rokhinson2012,Wiedenmann2016,Bocquillon2017,Wang2018,Peter2019,Calvez2019,Rosenbach2021,Mengmeng2022,Matthias2023}). In general, we show that an accurate theoretical description of driven (topological) quantum systems needs the careful treatment of the dependence of the basis on the external control parameter.

\begin{figure}[t]
	\centering
	\includegraphics[scale=1]{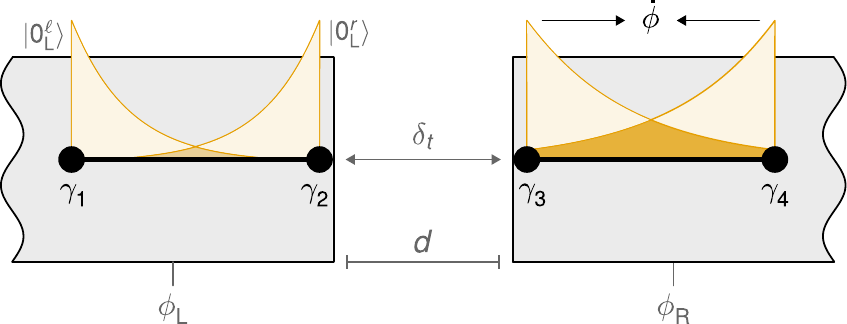}
	\caption{Two tunnel-coupled topological superconducting wires, represented by the four Majoranas~$\{ \gamma_1, \gamma_2, \gamma_3, \gamma_4 \}$. The wires are placed on top of two superconductors (\emph{gray}), which interconnect at the far ends to form a loop threaded by a time-dependent magnetic field. The applied field leads to a phase difference~$\phi = \phi_\text{R} - \phi_\text{L}$. The phase drop is included entirely in the right wire. The induced emf (owing to the time-dependent phase) enhances the coupling between Majoranas~$\gamma_3$ and~$\gamma_4$, as depicted by the overlap between the exponential tails of the left~$\ket{0^\ell_\alpha}$ and right~$\ket{0^r_\alpha}$ Majorana modes with~$\alpha \in \{\text{L},\text{R} \}$.}
	\label{Figure1}
\end{figure}

\emph{Model.}\textemdash We consider a flux-controlled Majorana junction, which consists of two tunnel-coupled topological superconductors. We model the superconductors as one-dimensional Kitaev chains with nearest-neighbor hopping and pairing~\cite{Kitaev2001,Leijnse2012}. The two contacts have a phase difference~$\phi \equiv \phi_\text{R} - \phi_\text{L}$ that is generally time dependent~(Fig.~\ref{Figure1}). The central question of this work is how to incorporate the time-dependent phase difference due to the bias voltage in the Hamiltonian description of the system. The most common choice is to attach~$\phi$ to the tunneling across the weak link, which leads to the Hamiltonian
\begin{equation} \label{ham_overall}
	H = H_\text{L} + H_\text{T}(\phi) + H_\text{R}\ .
\end{equation}
The individual Kitaev chains (of~$J$ sites) are described by
\begin{align}\label{Kitaev_Hamiltonian}
	H_{\alpha} = &-\mu \sum_{j}^{J} d_{j, \alpha}^\dagger d_{j, \alpha}^{} - t  \sum_{j}^{J- 1} \Big(  d^\dagger_{j+1, \alpha} d_{j, \alpha}^{} + \text{H.c.} \Big)  \nonumber \\	&+ \Delta  \sum_{j}^{J - 1} \Big( d_{j+1, \alpha}^{} d_{j,\alpha}^{}
	+ \text{H.c.} \Big),
\end{align}
where the chain index~$\alpha \in \{ \text{L, R} \}$, $t$ is the hopping amplitude, the chemical potential is within the topological limit~$|\mu| \le 2t$, and the pairing potential~$\Delta$ is real. The operator~$d_{j, \alpha}$ annihilates an electron at site~$j$ of chain~$\alpha$. The tunneling Hamiltonian reads
\begin{equation}\label{eq_H_T}
H_\text{T}(\phi) = - \delta t \, \left(  e^{-i\phi/2} d_{1, \text{R}}^\dagger d_{J,\text{L}}^{} +   e^{i\phi/2} d_{J, \text{L}}^\dagger d_{1, \text{R}}^{} \right),
\end{equation}
which couples the two chains with an amplitude~$\delta t \ll \ t $. 

The phase difference~$\phi$ could, however, be attached elsewhere in Hamiltonian or, in the most general case, be distributed along a given spatial profile within~$H$. This profile can be encoded in the basis choice of the Hamiltonian via a unitary transformation~$U(\phi)$~\cite{You2019,Riwar2021,Kenawy2022}. If~$\phi$ is constant in time, these choices are all gauge choices\textemdash that is, as long as Cooper pairs acquire the total phase~$\phi$ when traveling from one contact to the other, it does not matter where they acquire it along the way. But the situation is radically different for~$\phi\rightarrow \phi(t)$. Here, Hamiltonians with different basis choices provide different dynamics~\cite{You2019}. As pointed out in Ref.~\cite{Riwar2021}, different profiles of phase distributions correspond to different choices of the vector potential whose time derivative contributes to the (gauge-invariant) electric field. In other words, an arbitrary choice would not correctly account for the part of the electric field induced by the time-varying flux\textemdash that is, the emf.

Let us now focus on the asymmetric choice for which the phase~$\phi$ is attached to the right Kitaev chain. This choice is described by~$\mybar{H}=\mybar{H}_\text{L}+\mybar{H}_\text{T} +\mybar{H}_\text{R}(\phi)$ such that
\begin{align} \label{unitary}
	\mybar{H} =  U H U^\dagger = e^{i \phi G_\text{R}} \, H \, e^{-i \phi G_\text{R}}\ ,
\end{align}
with 
\begin{equation}\label{first_def_G}
	G_\text{R} = - \frac{1}{2} \sum_{j}^J  d_{j,\text{R}}^\dagger d_{j, \text{R}}\ .
\end{equation}
With this unitary transformation, the phase attaches to the pairing term in the right chain such that~$\Delta\rightarrow \Delta e^{\pm i\phi}$ for~$\mybar{H}_\text{R}$. As detailed in~\cite{Kenawy2022}, the phase distribution in~$\mybar{H}$ represents either a highly asymmetric placement of the chains with respect to the bulk superconductors (e.g., more phase drop is assigned to the right chain as it extends to the gap between the two superconductors) or a highly asymmetric distribution of the magnetic and induced electric fields. Physically, it corresponds to the case where the voltage drop~$V \equiv \dot{\phi}/(2e)$ occurs between the right chain and the right superconducting bulk. While more sophisticated phase profiles are by all means possible~\cite{Kenawy2022}, we take this choice as a simplified, generic representation of an asymmetric device geometry or asymmetrically applied fields. 

\emph{Low-energy Hamiltonian.}\textemdash We now derive a low-energy description for~$\mybar{H}$. The full derivation is given in the supplemental material, whereas here we summarize the main steps. First, it is useful to transform back into the basis choice that attaches the phase to the weak link (as in~$H$) to ensure that the low-energy basis is~$\phi$-independent. This transformation is accomplished by~$U$ as defined in Eqs.~\eqref{unitary} and~\eqref{first_def_G}. Due to the time-dependent basis change, the Schrödinger equation is now governed by~$ H +  \dot{\phi} G_\text{R}$, where the second term originates from the time derivative of the unitary transformation~$U$~\footnote{The unitary transformation~\eqref{first_def_G} leads to a~$4\pi$-periodic Hamiltonian, compared to the~$2\pi$-periodic one in Eq.~\eqref{ham_overall}. The change of periodicity does not, however, alter the size of the Hilbert space nor the fermion parity.}.

Next, we note that the second-quantized operators can be written in the form~$ A_\alpha = \thalf \psi^\dagger_\alpha \mathcal{A}_\alpha \psi_\alpha$, where the chain index~$\alpha \in \{\text{L},\text{R} \}$, the operator~$A_\alpha \in \{ H_\text{L}, H_\text{R}, G_\text{R} \}$, and the matrix~$\mathcal{A}_\alpha \in \{ \mathcal{H}_\text{L}, \mathcal{H}_\text{R}, \mathcal{G}_\text{R} \}$. The matrices~$\mathcal{H}_\alpha$ and~$\mathcal{G}_\text{R}$ are constructed such that their product with the field operators~$ \psi_\alpha = ( d^{}_{1, \alpha}, d^\dagger_{1, \alpha}  , \hdots  ,d^{}_{J, \alpha}, d^\dagger_{J, \alpha} )^\text{T}$ returns the many-body operators. Likewise, we can write the tunneling Hamiltonian as~$H_\text{T}(\phi) = \psi^\dagger_\text{L} \mathcal{H}_\text{T}(\phi) \psi_\text{R}$. 

The uncoupled chain Hamiltonians can be decomposed as~$\Ham_{\alpha} =  \Sigma_{v_\alpha} \epsilon_{v_\alpha} \ket{v_\alpha} \bra{v_\alpha}$. The particle-hole symmetry of Hamiltonian~$\mathcal{H}_\alpha$ implies that each eigenstate~$ \ket{v_\alpha}$ at energy~$\epsilon_{v_\alpha}$ has a pair~$\ket{\tilde{v}_\alpha}$ at energy~$-\epsilon_{v_\alpha}$, with the two related by~$\ket{v_\alpha} = \tau_x \ket{\tilde{v}_\alpha}$ where the Pauli matrix~$\tau_x$ acts on the Nambu space. The two Majoranas of each chain are related to the states~$\ket{0_\alpha}$ and~$\ket{\tilde{0}_\alpha}$. If the left and right Majorana modes do not overlap, these states are degenerate at zero energy, which is why we refer to them from now on as the zero-energy states (even though the degeneracy might in general be lifted).

For a low-energy description of the junction Hamiltonian, we eliminate all but the subspace comprising the zero-energy states~$\ket{0_\alpha}$ and~$\ket{\to_\alpha}$ of each chain, following a standard Schrieffer-Wolff transformation that includes higher-order corrections~\cite{Winkel}. This subspace is described by the projection operators~$\mathcal{P}_\alpha=\ket{0_\alpha}\bra{0_\alpha}+\ket{\to_\alpha}\bra{\to_\alpha}$, whereas~$\mathcal{Q}_\alpha = 1 - \mathcal{P}_\alpha$ projects onto the high-energy quasiparticle states.

After the projection, we obtain the Hamiltonian
\begin{equation}\label{eq_H_eff}
    	H_\text{low} = i \frac{\epsilon_0^{}}{2}   \gamma_2 \gamma_1 + i  \frac{ \epsilon_0^{} + g_\text{R}^{} }{2}   \gamma_4 \gamma_3  + i E_\text{M}  \cos \left( \frac{\phi}{2} \right) \gamma_2 \gamma_3,
\end{equation}
where, in leading order, the weak link only couples~$\gamma_2$ and~$\gamma_3$~(Fig.~\ref{Figure1}), with the operators~$\gamma$ denoting the second-quantized operators of the Majorana modes of each chain, which satisfy~$\{\gamma_\mu,\gamma_\nu\}=2i\delta_{\mu\nu}$. The energy~$\epsilon_0$ represents the previously mentioned ordinary overlap between left and right Majorana modes of each chain and equals~$ \bra{0_\alpha} \mathcal{H}_\alpha \ket{0_\alpha}$ (where we drop the chain index because we here assume the two chains to have identical parameters for simplicity). The low-energy projection of the extra term~$ \dot{\phi} {G}_\text{R}$ appears in the Hamiltonian~$H_\text{low}$ in two places: it modifies the Josephson energy~$E_\text{M}(\dot{\phi})$ and yields a new overlap term~$g_\text{R}(\dot{\phi})$. 

The former originates from the tunneling matrix~$\mathcal{H}_\text{T}(\phi)$ and couples the two Majoranas~$\gamma_2^{}$ and~$\gamma_3^{}$ across the junction. It is renormalized based on the time derivative of the phase~$\phi$. As discussed in the supplemental material, it is sufficient to account for emf-induced corrections to the Josephson energy~$E_\text{M}$ perturbatively in the form
\begin{equation}\label{scaled_josephson}
	E_\text{M}(\dot{\phi}) =  \bra{0_\text{L}} \mathcal{H}_\text{T}(0) \ket{0_\text{R}} -  \bra{0_\text{L}}   \mathcal{H}_\text{T}(0)  \, \frac{\mathcal{Q}_\text{R}}{ \mathcal{H}_\text{R} } \, \dot{\phi} {\mathcal{G}}_\text{R} \ket{0_\text{R}},
\end{equation}
where the first term is the standard fractional Josephson effect. The (first-order in~$\dot{\phi}$) correction term is the subject of our previous work~\cite{Kenawy2022}, where it was shown to be measurable either in the linear current response, or (in an open circuit geometry) as an additional contribution to charge fluctuations.

The new overlap term~$g_\text{R}$\textemdash similar to~$\epsilon_0$\textemdash couples the left and right Majorana modes of the right chain, but can be much larger in magnitude. As a matter of fact, it oscillates as a function of the voltage~$V = \dot{\phi}/(2e)$ and therefore, unlike~$E_\text{M}$, cannot be obtained using a perturbative expansion. We instead resum all higher-order corrections (while ignoring those that are exponentially suppressed in comparison~\footnote{Within the same order of~$\dot{\phi}$, we can compare two terms such as~$\mathcal{P}_\text{R} \mathcal{G}_\text{R} \mathcal{P}_\text{R} \mathcal{G}_\text{R} \mathcal{Q}_\text{R}  \mathcal{G}_\text{R} \mathcal{P}_\text{R}$ and~$\mathcal{P}_\text{R} \mathcal{G}_\text{R} \mathcal{Q_\text{R}} \mathcal{G}_\text{R} \mathcal{Q}_\text{R}  \mathcal{G}_\text{R}\mathcal{P}_\text{R}$. The former depends on the overlap~$\mathcal{P}_\text{R} \mathcal{G}_\text{R}\mathcal{P}_\text{R}$ between the exponentially decaying tails of the left and right Majoranas of the right chain, while the latter avoids this suppression and couples solely via the excited states described by the projector~$\mathcal{Q}_\text{R}$. The former is therefore much smaller and can be dropped. We follow this pattern to obtain the resummation~\eqref{resum}.}) to obtain (see the supplemental material)
\begin{equation}\label{resum}
	g_\text{R} (\dot{\phi}) = \bra{0_\text{R}} \left( \mathbb{\mathds{1}} + \dot{\phi} {\mathcal{G}}_\text{R} \frac{\mathcal{Q}_\text{R}}{\mathcal{H}_\text{R} - \epsilon_0} \right)^{-1}  \dot{\phi} {\mathcal{G}}_\text{R} \ket{0_\text{R}}.
\end{equation}
As we show below, in a relatively large parameter regime, this term can dominate both~$\epsilon_0$ and~$E_\text{M}$ and therefore significantly change the dynamics and the transport behavior of the junction.

On a fundamental level, it affects the symmetries present in the low-energy subspace. In the absence of the drive (i.e.,~$\dot{\phi} = 0$), if~$\epsilon_0 = 0$ (negligible overlap), the resulting chiral symmetry $\Gamma H_\text{low}\Gamma^\dagger = - H_\text{low}$ (where we can choose either~$\Gamma=i\gamma_1\gamma_2$ or~$\Gamma=i\gamma_3\gamma_4$ without loss of generality) leads to a gapless spectrum when coupling the two chains via the Josephson energy~$E_\text{M}$. Conversely, the finite overlap between Majorana edge modes yields~$\epsilon_0 \neq 0$, breaks chiral symmetry, and leads to finite gap in the spectrum of the coupled chains. But, for sufficiently long Kitaev chains, it can be expected that~$\epsilon_0$ is exponentially suppressed. The same, however, does not apply in general for~$g_\text{R}$. The emf produces a nonlinear overlap term that lifts the ground-state degeneracy of the right Kitaev chain, thereby breaking the chiral symmetry. As a simplistic intuitive picture, one can interpret the time-dependent driving of the phase as an effective reduction of the pairing potential~$\Delta$, which increases the overlap of the edge modes (whose coherence length is inversely proportional to~$\Delta$).

\begin{figure}[t]
	\centering
	\includegraphics[scale=1]{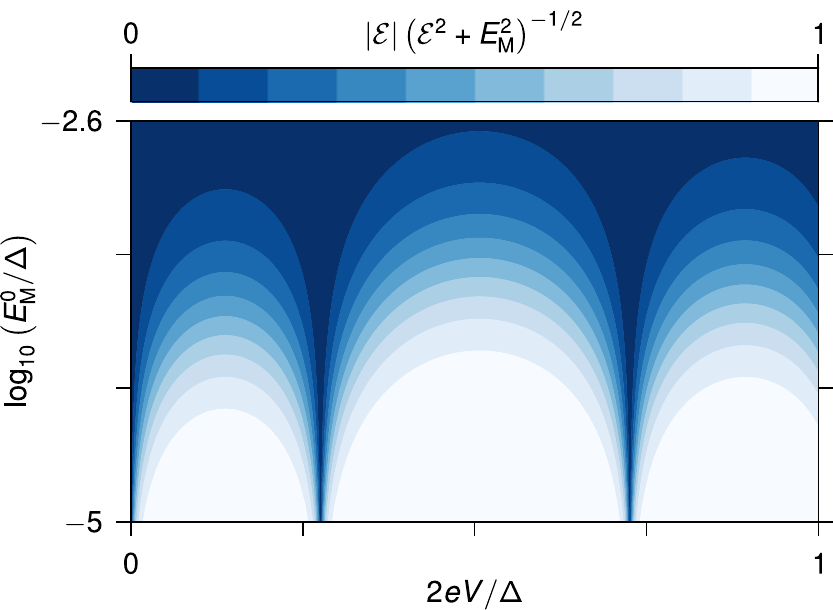}
	\caption{Energy spectrum of Hamiltonian~\eqref{subspace_ham} with odd parity as a function of the voltage~$V = \dot{\phi}/(2e)$ and the unrenormalized Josephson energy~$E_\text{M}^0 = E_\text{M}(\dot{\phi} = 0)$. We characterize the spectrum by the ratio~$|\mathcal{E}| (\mathcal{E}^2 + E_\text{M}^2)^{-1/2}$ where~$\mathcal{E} = g_\text{R} /2$ for odd parity~$(p=1)$ (i.e., in the absence of the drive, the uncoupled Kitaev chains have ground-state degeneracy at zero energy). The overlap energy~$g_\text{R}$ oscillates as a function of~$V$, most notably at smaller tunneling amplitudes. Parameters:~$J= 200$,~$\Delta = 0.04t$, and~$\mu = -1.95 t$.}
	\label{Figure2}
\end{figure}

\emph{Results.}\textemdash To understand how the emf-induced renormalization of parameters affects transport across the driven junction, we analyze the qubit formed by the coupling of the Majoranas of the left and right chains. We decompose the low-energy Hamiltonian into two uncoupled two-level systems for odd and even electron parities (see the supplemental material). The Hamiltonian of these two subspaces reads
\begin{equation}\label{subspace_ham}
	H_p = -\mathcal{E} \sigma_x + E_\text{M}  \cos \left( \frac{\phi}{2} \right) \sigma_z,
\end{equation}
where~$\mathcal{E} \equiv \delta_{p,0} \, \epsilon_0 + g_\text{R} / 2$ with the integer~$p = 0, 1$ for the even and odd parities, respectively. Let us focus on applying a constant voltage~($V = 2e\dot{\phi}=\text{constant})$. We can then understand the effect of the driving as a constant voltage-dependent renormalization of the energy scales~$\mathcal{E}$ and~$E_\text{M}$. The phase~$\phi$ inside the cosine remains the only time-dependent parameter. Consequently, the impact of a large~$g_\text{R}$ can be regarded as strong gapping of the instantaneous energy spectrum of Hamiltonian~$H_p$. Let us focus on the odd parity~$( p = 1)$ for which the spectrum is gapless in the absence of the drive~$(\dot{\phi} = 0)$. Importantly, the emf-induced gap oscillates as a function of the voltage~$V$. The smaller the Josephson energy~$ E_\text{M}^0$ in the absence of the drive (i.e., the worse the quality of the tunnel junction), the more pronounced the oscillations because the spectrum flattens\textemdash that is, the overlap energy~$g_\text{R}$ dominates the Josephson energy~$E_\text{M}$. A strong gapping furthermore implies a nearly flat instantaneous eigenspectrum as a function of~$\phi$, leading to a strongly suppressed supercurrent. To quantify the flatness, we plot the ratio~$\mathcal{E}/\sqrt{\mathcal{E}^2+E_\text{M}^2}$ in Fig.~\ref{Figure2}.

For a realistic description of the time evolution, we include a generic dissipative process captured by the Lindblad master equation
\begin{equation}\label{lindblad}
	\dot{\rho} = -i \left[ {H}_p, \rho \right] +  \Gamma \, D[L] \rho, 
\end{equation}
where the superoperator~$D[L]\rho \equiv L \rho L^\dagger - (1/2) \{ L^\dagger L, \rho\}$ represents the relaxation processes and, for simplicity, we only consider parity-conserving processes~\cite{Pekola2019}~\footnote{At least for topologically trivial junctions, experimental evidence suggests that parity flips can be made rare~\cite{Pekola2019}. Parity-conserving relaxation processes are therefore sufficient for qualitatively realistic predictions since the Hamiltonian~$H_p$ depends weakly on~$p$.}. It is convenient to work in the instantaneous eigenbasis of Hamiltonian~\eqref{subspace_ham}, defined by the unitary transformation
\begin{equation}\label{rotation}
	R = \frac{1}{\sqrt{2}} \begin{pmatrix}
	    \beta_- & \beta_+  \\
   \text{sgn} \left( \mathcal{E} \right) \, \beta_+  & -\text{sgn} \left( \mathcal{E} \right) \, \beta_-
	\end{pmatrix} ,
\end{equation}
where~$\beta_\pm = [ 1 \pm \lambda^{-1} E_\text{M} \cos(\phi/2) ]^{1/2}$ and the absolute value of the eigenvalue of~$H_p$ is~$\lambda = [\mathcal{E}^2 + E_\text{M}^2  \cos^2(\phi/2)]^{1/2}$. In this eigenbasis, the time evolution becomes governed by the Hamiltonian~$R^\dagger H_p R - i R^\dagger \dot{R} = - \lambda \, \sigma_z + Y \sigma_y$ with
\begin{equation}\label{coupl}
	Y \equiv - \dot{\phi} \, \frac{  |\mathcal{E}|  E_\text{M}}{4 \lambda^2 }  \, \sin \left(  \frac{\phi}{2} \right).
\end{equation}

For the dissipative term, we consider a single jump operator that relaxes the two-level system to its instantaneous ground state with a rate~$\Gamma$~(i.e.,~$L = \sigma_-)$. This process is consistent with the assumption that the bath is at zero temperature, and its correlation time is much shorter than~$(1/\mathcal{E})$ and~$ (1/E_\text{M})$. Defining the column vector~$ \ket{\tilde{\rho}}\rangle = \left( \tilde{\rho}_{00}, \tilde{\rho}_{01}, \tilde{\rho}_{10}, \tilde{\rho}_{11} \right)^\text{T} $, with~$\tilde{\rho}$ being the density matrix in the instantaneous eigenbasis, leads to the time-evolution equation~$	\ket{\dot{\tilde{\rho}}}\rangle  = \mathcal{L} \ket{\tilde{\rho}}\rangle $ with the Liouvillian
\begin{equation}
	 \mathcal{L}(t) = \begin{pmatrix}
	 	0& -Y & -Y & \Gamma \\
	 	Y& 2i\lambda-\frac{\Gamma}{2} & 0 & -Y \\
	 	Y & 0 & -2i\lambda-\frac{\Gamma}{2} & - Y \\
	 	0 & Y & Y & -\Gamma
	 \end{pmatrix}.
\end{equation} 

Here, we are interested in how the emf-induced renormalization of the parameters modifies the~$IV$ characteristics of the driven junction. The expected value of the current is defined as~$\text{Tr} ( \tilde{\rho} 	{\tilde{I}}  )$, where the current operator~$\tilde{I}$ in the instantaneous eigenbasis equals~$R^\dagger {I} R$ with
\begin{equation}
	I = e E_\text{M} \, \sin \left(\frac{\phi}{2}\right) \sigma_z.
\end{equation}
In a transient state (when the system did not have time to relax) the current does not need to have any particular periodicity in time. In the steady state, however, the current defaults to~$2\pi$ periodicity because the system is given sufficient time to mix between the two available states. Here, we focus on the dc current~$I_\text{dc}$, compared to the current~$I_\text{dc}^0$ without renormalization (i.e., substituting with~$E_\text{M}^0 \equiv E_\text{M}(\dot{\phi} = 0)$ and~$g_\text{R}(\dot{\phi} = 0)=0$ in~$H_p$). The unrenormalized parameters correspond to assigning the entire time-dependent phase difference to the weak link as in Eq.~\eqref{ham_overall}\textemdash the default assumption previous to our work.

\begin{figure}[t]
	\centering
	\includegraphics[scale=1]{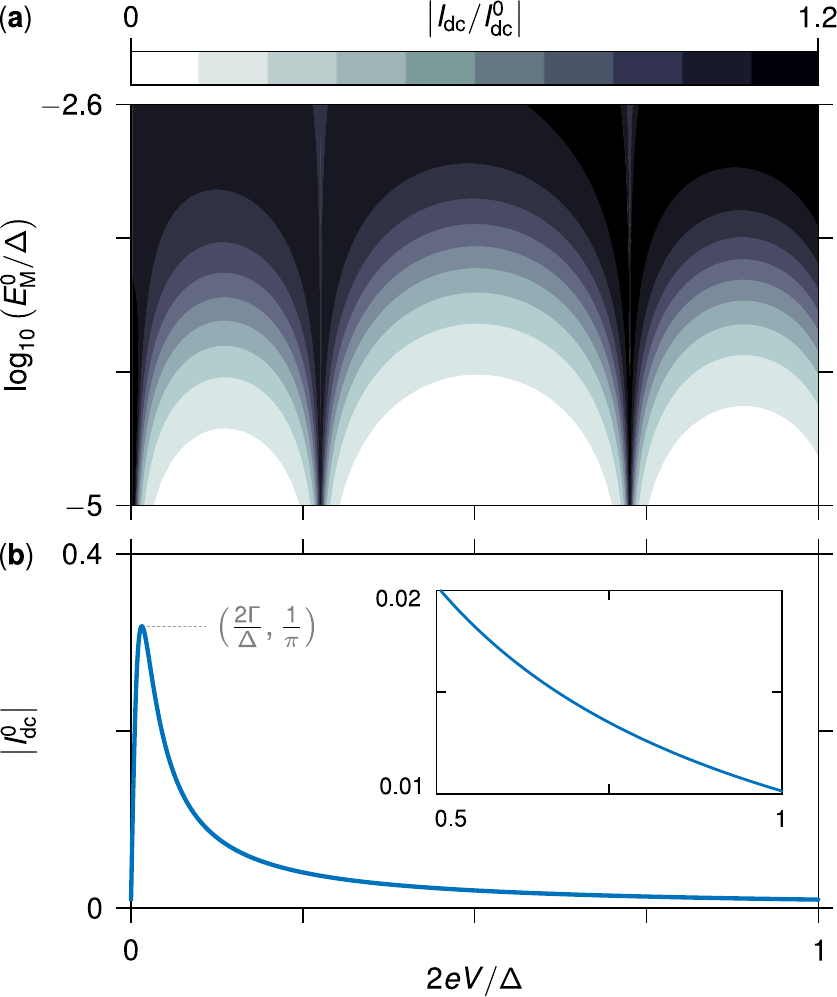}
	\caption{Steady-state dc current across the junction. (\textbf{a})~The dc current~$I_\text{dc}$ is normalized by~$I_\text{dc}^0$ obtained using the unrenormalized energies~$E_\text{M}^0 \equiv E_\text{M}(\dot{\phi} = 0)$ and~$g_\text{R}(\dot{\phi} = 0) = 0$, which correspond to including the entire time-dependent phase~$\phi$ in the weak link. The dc current~$I_\text{dc}$ is strongly suppressed, especially for smaller Josephson energies (i.e.,~at weaker tunneling between the two Kitaev chains). It also oscillates as a function of the voltage~$V = \dot{\phi}/(2e)$, exhibiting the same structure as the energy spectrum in Fig.~\ref{Figure2}. (\textbf{b})~The dc current~$I_\text{dc}^0$ is normalized by the critical current~$e |{E}^0_\text{M}|$. The chain parameters are the same as Fig.~\ref{Figure2}. The relaxation rate~$\Gamma = \Delta / (40\pi)$ so that, in the range of voltages considered, the dynamics are not dominated by relaxation.}
	\label{Figure3}
\end{figure}

Let us again discuss the odd parity~$p=1$. Without renormalizing the parameters, the spectrum of the driven junction is still gapless. In this case, the steady-state dc current has the analytical solution (see the supplemental material)
\begin{align}
    I_\text{dc}^0 = -\frac{1}{\pi} \frac{4 \Gamma  \dot{\phi} }{4 \Gamma^2 +  \dot{\phi}^2  },
\end{align}
which is in units of the critical current~$e|E_\text{M}^0|$. The dependence of the current~$I_\text{dc}^0$ on the voltage~$V = \dot{\phi}/(2e)$ is depicted in Fig.~\ref{Figure3}(\textbf{b}).

Using the renormalized parameters, on the other hand, reveals that the time-dependent phase~$\dot{\phi}$ results in a much richer physics for the driven system, as captured by Fig.~\ref{Figure3}(\textbf{a}). The dc current in the steady state exhibits the same structure as the spectrum in Fig.~\ref{Figure2}, reflecting the fact that the flattening of the instantaneous spectrum indeed suppresses the current across the junction. Moreover, the smaller the Josephson energy~$E_\text{M}^0$ (i.e., the worse the quality of the tunnel junction), the stronger the effect of renormalizing the parameters. This behavior demonstrates one of our main results: the initially linear emf correction term~$-i\dot{\phi} U^\dagger \partial_\phi U$ can yield a strongly nonlinear effect on the system dynamics when projected to the low-energy subspace.

\emph{Conclusion.}\textemdash We study the interplay between topological protection and a classical time-dependent driving through the representative example of Majorana junctions. By deriving a low-energy theory, we show that the induced electromotive force (emf) modifies the Josephson energy and enhances the effective overlap between the left and right Majorana modes of the nanowires. The renormalization of these two energy scales manifests as a strong suppression in the steady-state dc current across the junction. Our work therefore illustrates the importance of a proper microscopic description of the coupling between a given quantum system and the control parameter. Finally, we note that the central ingredients for the physics discussed in this paper (time-dependent control in the presence of a time-dependent basis and the exponential suppression of topological edge-state overlap) transcend the narrow context of Majorana systems. It remains therefore an interesting topic for future research to extend the analysis presented in this paper to the protection of other topological systems.

\bibliography{refLib}

\newpage\hbox{}\thispagestyle{empty}\newpage

\widetext
\begin{center}
\textbf{\large Supplemental Material: Time-dependent driving and topological protection in the fractional Josephson effect}
\end{center}
%%%%%%%%%% Merge with supplemental materials %%%%%%%%%%
%%%%%%%%%% Prefix a "S" to all equations, figures, tables and reset the counter %%%%%%%%%%
\setcounter{equation}{0}
\setcounter{figure}{0}
\setcounter{table}{0}
\setcounter{page}{1}
\makeatletter
\renewcommand{\theequation}{S\arabic{equation}}
\renewcommand{\thefigure}{S\arabic{figure}}
\renewcommand{\bibnumfmt}[1]{[S#1]}
\renewcommand{\citenumfont}[1]{S#1}

\section{Low-energy description for the driven junction}

In this section of the supplemental material, we derive the low-energy Hamiltonian~$H_\text{low}$ in Eq.~\eqref{eq_H_eff} in the main text. In the basis that attaches the entire phase difference (between the two superconductors in Fig.~1) to the weak link, the time-dependent Schrödinger equation is governed by~$H + \dot{\phi} G_\text{R}$, where~$H$ is given in Eqs.~\eqref{ham_overall},~\eqref{Kitaev_Hamiltonian}, and \eqref{eq_H_T}, whereas~$G_\text{R}$ is defined in Eq.~\eqref{first_def_G} in the main text.

In the Bogoliubov–de Gennes (BdG) form, the Hamiltonian can be written as
\begin{equation} 
	\begin{split}
		H &= \half  \psi^\dagger \mathcal{H} \psi \\ &=\half    \begin{pmatrix}
			\psi^\dagger_\text{L} & 	\psi^\dagger_\text{R}  
		\end{pmatrix} 
		\begin{pmatrix}
			\Ham_{\text{L}}  &\mathcal{H}_\text{T}(\phi) \\   \mathcal{H}_\text{T}^\dagger(\phi) &  \Ham_{ \text{R}} 
		\end{pmatrix}
		\begin{pmatrix}
			\psi^{}_\text{L} \\	\psi^{}_\text{R}
		\end{pmatrix} ,
  \end{split}
\end{equation}
where~$ \psi_\alpha = ( d^{}_{1, \alpha}, d^\dagger_{1, \alpha}  , \hdots  ,d^{}_{J, \alpha}, d^\dagger_{J, \alpha} )^\text{T}$. In terms of its eigenstates, the uncoupled chain Hamiltonians can be decomposed as
\begin{equation}
    \Ham_{\alpha} =  \sum_{v_\alpha} \epsilon_v \ket{v_\alpha} \bra{v_\alpha} - \epsilon_v \ket{\tilde{v}_\alpha} \bra{\tilde{v}_\alpha}.
\end{equation}
Each eigenstate~$ \ket{v_\alpha}$ at energy~$\epsilon_{v_\alpha}$ has a pair~$\ket{\tilde{v}_\alpha}$ at energy~$-\epsilon_{v_\alpha}$, with the two related by~$\ket{v_\alpha} = \tau_x \ket{\tilde{v}_\alpha}$ where the Pauli matrix~$\tau_x$ acts on the particle and hole blocks. The two Majoranas of each chain are related to the states~$\ket{0_\alpha}$ and~$\ket{\tilde{0}_\alpha}$, which are degenerate at zero energy if the left and right Majorana modes do not overlap. The tunneling matrix can be written as
\begin{equation} 
		\mathcal{H}_\text{T}^{j,j\prime}(\phi) = \delta_{j,J} \,  \delta_{j^\prime, 1} 
		\begin{pmatrix}
			- \delta_t \, e^{-i\phi/2} & 0  \\    0 &  \delta_t \, e^{i\phi/2} 
		\end{pmatrix}.
\end{equation}
Finally, 
the operator~$G_\text{R}$ reads
\begin{align}
        G_\text{R} = \frac{1}{2} \psi^\dagger \begin{pmatrix}
			0 & 0  \\    0 &  \mathcal{G}_\text{R}
		\end{pmatrix} \psi,
\end{align}
where~$\mathcal{G}_\text{R} = -\tau_z/2$. %because the original phase distribution in Hamiltonian~(2) in the manuscript includes the entire phase drop as a constant along the right chain.

Our goal is to obtain a Hamiltonian that describes the low-energy subspace defined by the projector
\begin{align} \label{projector_p}
	\mathcal{P}  &=  \begin{pmatrix}
		\mathcal{P}_\text{L} & 0 \\
		0 & 	\mathcal{P}_\text{R}
	\end{pmatrix}\nonumber \\ &= \sum_{v = 0, \tilde{0}}  \begin{pmatrix}
 \ket{v_\text{L} } \bra{v_\text{L}} & 0 \\
	0 &  \ket{v_\text{R} } \bra{v_\text{R}}
\end{pmatrix}.
\end{align}
We derive the low-energy Hamiltonian by decoupling the subspace comprising the four states~$\{ \ket{0_\text{L}}, \ket{\to_\text{L}}, \ket{0_\text{R}}, \ket{\to_\text{R}} \} $ from the high-energy states defined by~$\mathcal{Q} = 1 - \mathcal{P}$ via a perturbative expansion in the tunneling amplitude~$\delta t$ and~$\dot{\phi}$. We focus on processes that are first order in~$\delta_t$ but account for higher orders in~$\dot{\phi}$. The general form of the low-energy Hamiltonian can be written as
\begin{equation}
    {H}_\text{low} = {H}_\text{low}^{(0)} + {H}_\text{low}^{(1)}   + {H}_\text{low}^{(2)}  + \dots,
\end{equation}
where the order of~${H}_\text{low}^{n}$ denotes the sum of the orders of both~$\delta_t$ and~$\dot{\phi}$. The form of the Schrieffer-Wolff transformation can be found in~\cite{Winkel}. For the zeroth order in~$\delta_t$ and~$\dot{\phi}$, we have
\begin{align}
    {H}_\text{low}^{(0)} &= \frac{1}{2} \sum_\alpha \psi^\dagger_\alpha  \left( \mathcal{P}_\alpha \mathcal{H}_\alpha \mathcal{P}_\alpha  \right) \psi_\alpha \nonumber \\ 
    &= \frac{1}{2} \sum_\alpha \psi^\dagger_\alpha  \left( \sum_{x_\alpha,y_\alpha} h_{xy}^\alpha \ket{x_\alpha}\bra{y_\alpha} \right) \psi_\alpha ,
\end{align}
where~$\{ x,y\} \in \{0,\tilde{0} \}$ and~$h^\alpha_{xy} = \bra{x_\alpha} \mathcal{H_\alpha} \ket{y_\alpha}$. The matrix element reads
\begin{equation}
    \bra{0_\alpha} \mathcal{H}_\alpha \ket{0_\alpha} = - \bra{\to_\alpha} \mathcal{H}_\alpha \ket{\to_\alpha} = \epsilon_0,
\end{equation}
where the subscript~$\alpha$ is dropped for the energy~$\epsilon_0$ because we assume that the two chains have the same parameters\textemdash namely,~$t,\mu$, and~$\Delta$. The zeroth-order term then simplifies to
\begin{align}\label{epsilon}
    {H}_\text{low}^{(0)} &= \frac{\epsilon_0}{2} \sum_\alpha \psi^\dagger_\alpha \left(  \ket{0_\alpha}\bra{0_\alpha} - \ket{\to_\alpha}\bra{\to_\alpha} \right) \psi_\alpha \nonumber \\ 
     &= \frac{\epsilon_0}{2} \sum_\alpha \left(c^\dagger_\alpha c_\alpha - c_\alpha c_\alpha^\dagger   \right) \nonumber \\
     &= \frac{\epsilon_0}{2} \sum_\alpha \left( 2c^\dagger_\alpha c_\alpha - 1  \right) \nonumber \\
     &= i \frac{\epsilon_0}{2} \gamma_4 \gamma_3 + i \frac{\epsilon_0}{2} \gamma_2 \gamma_1,
\end{align}
where the Majorana operators are defined via
\begin{equation}\label{cLL}
	c_\text{L}^{} = \frac{\gamma_2 + i \gamma_1}{2},
\end{equation}
and
\begin{equation}\label{cRR}
	c_\text{R}^{} = \frac{\gamma_4 + i \gamma_3}{2}.
\end{equation}

Next, we focus on terms that are zeroth order in~$\delta_t$ but nonzero in~$\dot{\phi}$. In analogy to Eq.~\eqref{epsilon}, for the right chain, we obtain the overlap energies
\begin{equation}
    \frac{i}{2} \left(  g_\text{R}^{(1)} + g_\text{R}^{(2)} + g_\text{R}^{(3)} + \dots \right) \, \gamma_4 \gamma_3,
\end{equation}
where the first-order term reads
\begin{equation}
    g_\text{R}^{(1)} = \bra{0_\text{R}}  \bar{\mathcal{G}}_\text{R} \ket{0_\text{R}},
\end{equation}
with~$\bar{\mathcal{G}}_\text{R} \equiv \dot{\phi} \mathcal{G}_\text{R}$. Similarly, the higher-order contributions read
\begin{equation}
    g_\text{R}^{(2)} = - \Bra{0_\text{R}} \bar{\mathcal{G}}_\text{R} \frac{\mathcal{Q}_\text{R}}{\mathcal{H}_\text{R} - \epsilon_0} \bar{\mathcal{G}}_\text{R} \Ket{0_\text{R}},
\end{equation}
and
\begin{equation}
    g_\text{R}^{(3)} =  \bra{0_\text{R}} \bar{\mathcal{G}}_\text{R} \frac{\mathcal{Q}_\text{R}}{\mathcal{H}_\text{R} - \epsilon_0} \bar{\mathcal{G}}_\text{R}  \frac{\mathcal{Q}_\text{R}}{\mathcal{H}_\text{R} - \epsilon_0} \bar{\mathcal{G}}_\text{R}\ket{0_\text{R}} - \bra{0_\text{R}} \bar{\mathcal{G}}_\text{R} \ket{0_\text{R}} \bra{0_\text{R}} \bar{\mathcal{G}}_\text{R} \frac{\mathcal{Q}_\text{R}}{\mathcal{H}_\text{R} - \epsilon_0} \frac{\mathcal{Q}_\text{R}}{\mathcal{H}_\text{R} - \epsilon_0} \bar{\mathcal{G}}_\text{R}\ket{0_\text{R}}.
\end{equation}
Importantly, the second term in~$g_\text{R}^{(3)}$ is exponentially suppressed compared to the first because it depends on the overlap between the exponentially decaying tails of the left and right Majorana modes through the matrix element~$\bra{0_\text{R}} \bar{\mathcal{G}}_\text{R} \ket{0_\text{R}}$. Conversely, the first term avoids this suppression by coupling exclusively via the excited states, denoted by the projector~$\mathcal{Q}_\text{R}$. The third-order term then reduces to
\begin{equation}
    g_\text{R}^{(3)} \approx  \bra{0_\text{R}} \bar{\mathcal{G}}_\text{R} \frac{\mathcal{Q}_\text{R}}{\mathcal{H}_\text{R} - \epsilon_0} \bar{\mathcal{G}}_\text{R}  \frac{\mathcal{Q}_\text{R}}{\mathcal{H}_\text{R} - \epsilon_0} \bar{\mathcal{G}}_\text{R}\ket{0_\text{R}}.
\end{equation}
This logic can be extended to arbitrary orders in~$\dot{\phi}$. We can therefore obtain a good approximation by partially resumming, that is, summing only terms of the first type (without exponential suppression) for each order in~$\dot{\phi}$. We find
\begin{align}
         g_\text{R}(\dot{\phi}) &=   g_\text{R}^{(1)} + g_\text{R}^{(2)} + g_\text{R}^{(3)} + \dots \nonumber \\
        &=  \bra{0_\text{R}} \left[ \bar{\mathcal{G}}_\text{R} + \bar{\mathcal{G}}_\text{R} \left(  - \frac{\mathcal{Q}_\text{R}}{\mathcal{H}_\text{R} - \epsilon_0} \bar{\mathcal{G}}_\text{R} \right)  + \bar{\mathcal{G}}_\text{R}  \left(  - \frac{\mathcal{Q}_\text{R}}{\mathcal{H}_\text{R} - \epsilon_0} \bar{\mathcal{G}}_\text{R} \right) \left(  - \frac{\mathcal{Q}_\text{R}}{\mathcal{H}_\text{R} - \epsilon_0} \bar{\mathcal{G}}_\text{R} \right) + \dots   \right] \ket{0_\text{R}} \nonumber \\
    &= \bra{0_\text{R}} \left(  \mathds{1} + \bar{\mathcal{G}}_\text{R}  \frac{\mathcal{Q}_\text{R}}{\mathcal{H}_\text{R} - \epsilon_0} \right)^{-1} \bar{\mathcal{G}}_\text{R}       \ket{0_\text{R}}.
\end{align}
Accordingly, the~$\dot{\phi}$-dependent corrections to the overlap energy in the right chain take the form
\begin{equation}\label{g_all}
    i\frac{g_\text{R}}{2} \, \gamma_4 \gamma_3\ ,
\end{equation}
with~$g_\text{R}$ as given in Eq.~\eqref{resum} in the main text.

Finally, we focus on terms that are proportional to the tunneling amplitude~$\delta_t$, starting with those that are zeroth order in~$\dot{\phi}$, namely
\begin{align}\label{zeroth_tunneling}
     \frac{1}{2} \left[   \psi^\dagger_\text{L}  ( \mathcal{P}_\text{L} \mathcal{H}_\text{T}(\phi)  \mathcal{P}_\text{R}  ) \psi_\text{R}  + \psi^\dagger_\text{R}  ( \mathcal{P}_\text{R} \mathcal{H}_\text{T}^\dagger(\phi) \mathcal{P}_\text{L}  ) \psi_\text{L}  \right].
\end{align}
The first term can be simplified to
\begin{equation}
        \mathcal{P}_\text{L} \mathcal{H}_\text{T} \mathcal{P}_\text{R} = \sum_{x,y = 0,\to} q_{xy}^{(1)} \ket{x_\text{L}} \bra{y_\text{R}},
\end{equation}
with
\begin{equation}
    q_{xy}^{(1)}(\phi) = \bra{x_\text{L}} \mathcal{H}_\text{T}(\phi) \ket{y_\text{R}}.
\end{equation}
Substituting with~$\mathcal{P}_\text{L} \mathcal{H}_\text{T} \mathcal{P}_\text{R}$ into Eq.~\eqref{zeroth_tunneling} leads to
\begin{align}\label{fermionic_zerothCoupling}
     &\frac{1}{2}   \left( q_{00}^{(1)} c_\text{L}^\dagger c_\text{R} + q_{\to 0}^{(1)} c_\text{L} c_\text{R} + q_{0\to}^{(1)} c_\text{L}^\dagger c_\text{R}^\dagger + q_{\to \to}^{(1)} c_\text{L} c_\text{R}^\dagger +\text{H.c.}  \right) \nonumber \\
    &= q_{00}^{(1)} c_\text{L}^\dagger c_\text{R} - q_{\to \to}^{(1)} c_\text{R}^\dagger c_\text{L} + q_{\to 0}^{(1)} c_\text{L} c_\text{R} - q_{0 \to}^{(1)} c_\text{R}^\dagger c_\text{L}^\dagger\ ,
\end{align}
where~$c_\text{L}$ and~$c_\text{R}$ are defined in Eqs.~\eqref{cLL} and~\eqref{cRR}, respectively. Alternatively, we can express the result in terms of the Majorana operators. In this case, Eq.~\eqref{fermionic_zerothCoupling} takes the form
\begin{equation}
         i \, \Re\left\{ \frac{q_{00}^{(1)} - q_{0 \to}^{(1)}}{2} \right\}  \gamma_2 \gamma_3 =
     i \left( \frac{q_{00}^{(1)}(0) - q_{0 \to}^{(1)}(0) }{2} \right) \, \cos\left(\frac{\phi}{2} \right) \,  \gamma_2 \gamma_3,
\end{equation}
where, in leading order, the weak link only couples Majoranas~$\gamma_2$ and~$\gamma_3$.

As for the corrections of the electromotive force (emf) to tunneling across the junction, it suffices to only include corrections that are first order in~$\dot{\phi}$, leading to
\begin{align}
    i \, \Re\left\{ \frac{q_{00}^{(2)} - q_{0 \to}^{(2)}}{2} \right\}  \gamma_2 \gamma_3,
\end{align}
where
\begin{equation}
    q_{xy}^{(2)}(\phi) = - \frac{1}{2} \bra{x_\text{L}}  \mathcal{H}_\text{T} (\phi) \frac{\mathcal{Q}_\text{R}}{\mathcal{H}_\text{R} - \epsilon_x} \bar{\mathcal{G}}_\text{R} + \mathcal{H}_\text{T}(\phi) \frac{\mathcal{Q}_\text{R}}{\mathcal{H}_\text{R} - \epsilon_y} \bar{\mathcal{G}}_\text{R} \ket{y_\text{R}}.
\end{equation}
The tunneling terms therefore become
\begin{equation}\label{josephson_qform}
    i \left( \frac{q_{00}(0) - q_{0 \to}(0) }{2} \right) \, \cos\left(\frac{\phi}{2} \right) \,  \gamma_2 \gamma_3,
\end{equation}
with
\begin{equation}
    q_{xy}(\phi) = q_{xy}^{(1)}(\phi) + q_{xy}^{(2)}(\phi) .
\end{equation}
In limit where only coupling between~$\gamma_2$ and~$\gamma_3$ across the weak link survives, we can simplify the coefficients to
\begin{equation}
    q_{00}^{(1)}(0) \approx - q_{0\to}^{(1)}(0) \approx \bra{0_\text{R}} \mathcal{H}_\text{T}(0) \ket{0_\text{R}},
\end{equation}
and
\begin{equation}
    q_{00}^{(2)}(0) \approx - q_{0\to}^{(2)}(0) \approx -  \bra{0_\text{L}}  \mathcal{H}_\text{T}(0) \frac{\mathcal{Q}_\text{R}}{\mathcal{H}_\text{R}} \bar{\mathcal{G}}_\text{R}  \ket{0_\text{R}} .
\end{equation}
We can therefore rewrite Eq.~\eqref{josephson_qform} as
\begin{equation}
    i  E_\text{M} \, \cos \left( \frac{\phi}{2} \right) \, \gamma_2 \gamma_3,
\end{equation}
with
\begin{equation}\label{EM_term}
    E_\text{M} (\dot{\phi}) = \bra{0_\text{R}} \mathcal{H}_\text{T}(0) \ket{0_\text{R}} -\bra{0_\text{L}}  \mathcal{H}_\text{T}(0) \frac{\mathcal{Q}_\text{R}}{\mathcal{H}_\text{R}} \bar{\mathcal{G}}_\text{R}  \ket{0_\text{R}}.
\end{equation}
Adding Eqs.~\eqref{epsilon},~\eqref{g_all}, and~\eqref{EM_term} leads to the low-energy Hamiltonian
\begin{equation}
    H_\text{low} = i \frac{\epsilon_0}{2} \gamma_2 \gamma_1 + i \left( \frac{\epsilon_0 + g_\text{R} }{2} \right) \gamma_4 \gamma_3 +  i  E_\text{M} \, \cos \left( \frac{\phi}{2} \right) \, \gamma_2 \gamma_3.
\end{equation}
Using Eq.~\eqref{cLL}, we can decompose the low-energy Hamiltonian into two decoupled two-level systems for odd and even electron parities. The odd parity consists of the states~$\ket{01}$ and~$\ket{10}$, while the even parity of~$\ket{00}$ and~$\ket{11}$ (with~$c_\text{L}^\dagger c_\text{R}^\dagger \ket{00} = \ket{11}$, and Eqs.~\eqref{cLL} and~\eqref{cRR} define~$c_\text{L}$ and~$c_\text{R}$, respectively). The Hamiltonians of these two subspaces read
\begin{equation}\label{subspace_ham_supp}
	H_p = -\mathcal{E} \sigma_x + E_\text{M}  \cos \left( \frac{\phi}{2} \right) \sigma_z,
\end{equation}
where~$\mathcal{E} \equiv \delta_{p,0} \, \epsilon_0 + g_\text{R} / 2$ with the integer~$p = 0, 1$ for the even and odd parities, respectively.

\section{Steady-state dc current for the ungapped Hamiltonian}

\begin{figure}[t]
	\centering
	\includegraphics[scale=1]{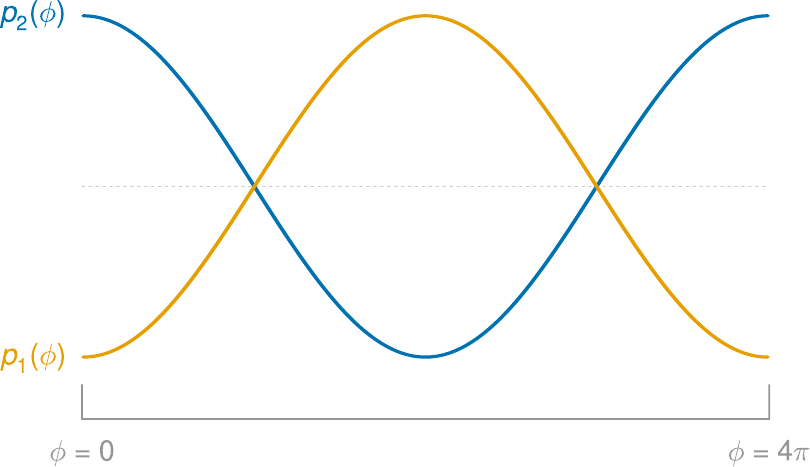}
	\caption{Energy spectrum of Hamiltonian~$H_p$ in Eq.~\eqref{subspace_ham} in the main text with~$\mathcal{E} = 0$. The zero gap occurs either due to the oscillatory behavior of~$g_\text{R}$ as a function of the voltage~$V$ (Fig.~\ref{Figure2} in the main text), or simply when using the unrenormalized parameters that assumes the entire phase drop between the two superconductors can be included across the weak link.}
	\label{Figure1S}
\end{figure}

This section derives an analytical expression for the dc current in the steady state when the energy spectrum is not gapped (i.e.,~$\mathcal{E}=0$ in Eq.~\eqref{subspace_ham_supp} or Eq.~\eqref{subspace_ham} in the main text). For a vanishingly small gap, the basis defined by the unitary transformation~\eqref{rotation} in the main text is not suitable since the off-diagonal term~\eqref{coupl} in the main text tends to a delta function. Instead, a suitable basis for the gapless system consists of the two uncoupled branches of~$E_\text{M} \cos(\phi/2) \, \sigma_z$ (Fig.~\ref{Figure1S}). In this basis, we can obtain an analytical solution for the steady-state occupation probabilities and, subsequently, the dc current. For a gapless system, the two eigenstates are not coupled and the only process that changes the occupation probabilities is relaxation. Assuming a constant voltage~$V$, it follows that~$\phi(t)=2eV t$ such that there is a one-to-one correspondence between time~$t$, and~$\phi$ at a given time. We therefore use the notation that when~$\phi$ appears as an argument for the time evolution of the state, it should be interpreted as the time at which the phase assumes the value of~$\phi$ (this avoids confusion with the tunneling parameter~$t$ in the Kitaev chain).

Starting from an initial value at~$\phi = 0$ (Fig.~\ref{Figure1S}), the probability at~$\phi = \pi$ can be written as
\begin{align}\label{pEq1}
    p_1(\pi) &= p_1(0) + \left(1 - W \right) \,p_2(0) \nonumber \\ 
    &= p_1(0) + (1 - W) \, \left[ 1 - p_1(0) \right],
\end{align}
where~$ W \equiv \text{exp} ( -\pi  \Gamma / \dot{\phi} )$. Here,~$p_1$ and~$p_2$ represent the occupation probabilities of the two states of the Hamiltonian~$E_\text{M} \cos(\phi/2) \, \sigma_z$ as defined in Fig.~\ref{Figure1S} (i.e., at~$\phi = 0$,~$p_1$ corresponds to the ground state and~$p_2$ to the excited state). Similarly, we can write the probability~$p_1$ at~$\phi = 3\pi$ as 
\begin{align}\label{pEq2}
    p_1(3\pi) = W^2 \, p_1(\pi ) ,
\end{align}
and, finally, at~$\phi = 4\pi$ as
\begin{align}\label{pEq3}
    p_1(4\pi) =  p_1(3\pi ) + (1 - W) \, \left[ 1 - p_1(3\pi) \right].
\end{align}
In the steady state, we can impose the~$4\pi$-periodicity condition
\begin{align}\label{Pcond}
    p_1(4\pi) =  p_1(0 ).
\end{align}
Solving the four equations~\eqref{pEq1},~\eqref{pEq2},~\eqref{pEq3}, and~\eqref{Pcond} yields the steady-state probabilities
\begin{align}
    p_1(0) = 1 - \frac{W}{1 + W^2},
\end{align}
and
\begin{align}
    p_2(0) =  \frac{W}{1 + W^2}.
\end{align}
The values~$\phi = 0$ can be readily used to obtain the probabilities as a function of~$\phi$. As for the current, its expected value is defined as~$ \text{Tr} ( \rho I )$, where the occupation probabilities are the diagonal elements of the density matrix~$ \rho$ and with the current operator
\begin{equation}
	I = e E_\text{M} \, \sin \left(\frac{\phi}{2}\right) \sigma_z.
\end{equation}
The expected value of the current then reads
\begin{align}
    I^\text{gapless} = e \left|E_\text{M} \right| \left[ 1 - 2 p_1(\phi) \right] \, \sin \left( \frac{\phi}{2} \right),
\end{align}
with the superscript denoting that this expression is valid for the gapless system. Averaging over the period~$\left[0, 4\pi\right ]$ yields the steady-state dc current
\begin{align}\label{dc_analytical}
    I_\text{dc}^\text{gapless} &=  \frac{e \left|E_\text{M} \right|}{4\pi} \int_0^{4\pi} I^\text{gapless} \, \mathrm{d}\phi \nonumber \\
    &= -\frac{e \left|E_\text{M} \right|}{\pi} \frac{4 \Gamma  \dot{\phi} }{4 \Gamma^2 +  \dot{\phi}^2  }.
\end{align}
The analytical solution~\eqref{dc_analytical} is valid at points where the new overlap energy~$g_\text{R}$ is zero (or vanishingly small) as a function of the voltage~$V$ (see Fig.~\ref{Figure2} and Fig.~\ref{Figure3}(\textbf{a}) in the main text). It is also used to obtain the dc current in Fig.~\ref{Figure3}(\textbf{b}) for the driven system without renormalizing the parameters (i.e.,~$g_\text{R} = 0$ and~$E_\text{M} \rightarrow  E_\text{M}^0 = E_\text{M}(\dot{\phi}= 0) )$. %This choice corresponds to assigning the entire phase difference to the weak link\textemdash the default assumption previous to our work.

\end{document}